\documentclass[12pt, twoside]{article}\usepackage[]{graphicx}\usepackage[]{color}
\makeatletter
\def\maxwidth{ %
  \ifdim\Gin@nat@width>\linewidth
    \linewidth
  \else
    \Gin@nat@width
  \fi
}
\makeatother

\definecolor{fgcolor}{rgb}{0.345, 0.345, 0.345}

\usepackage{framed}
\makeatletter
 {\par\unskip\endMakeFramed%
 \at@end@of@kframe}
\makeatother

\definecolor{shadecolor}{rgb}{.97, .97, .97}
\definecolor{messagecolor}{rgb}{0, 0, 0}
\definecolor{warningcolor}{rgb}{1, 0, 1}
\definecolor{errorcolor}{rgb}{1, 0, 0}
\newenvironment{knitrout}{}{} 

\usepackage{alltt}

\definecolor{contourblue}{rgb}{.235,.392,.49}

\usepackage[margin=1in]{geometry}

\usepackage[font=doublespacing]{caption}
\usepackage{setspace}

\usepackage{amsthm, amssymb, bm, dsfont, mathrsfs}
\usepackage[leqno]{amsmath}
\usepackage{chngcntr}
\usepackage{apptools}
\AtAppendix{\counterwithin{theorem}{subsection}}
\AtAppendix{\counterwithin{lemma}{subsection}}
\AtAppendix{\counterwithin{corollary}{subsection}}

\usepackage{booktabs, makecell, enumitem, tabularx, multirow}
\newcommand{\ra}[1]{\renewcommand{\arraystretch}{#1}}

\newcolumntype{Y}{>{\centering\arraybackslash}X}

\usepackage{chngcntr}

\usepackage{xparse}

\usepackage{xr}
\usepackage[natbibapa]{apacite}
\usepackage{etoolbox}

\AtBeginEnvironment{APACrefURL}{\renewcommand{\url}[1]{}}
\RequirePackage[colorlinks,linkcolor=blue,citecolor=blue,urlcolor=blue]{hyperref}
\usepackage{cleveref}
\crefname{lemma}{Lemma}{Lemmas}
\crefname{corollary}{Corollary}{Corollaries}

\binoppenalty=\maxdimen
\relpenalty=\maxdimen

\usepackage{placeins}
\usepackage[nolists, nomarkers, noheads, tablesfirst]{endfloat}

\def\equationautorefname~#1\null{%
  (#1)\null
}

\allowdisplaybreaks

\makeatletter
\newcommand\Autoref[1]{\@first@ref#1,@}
\def\@throw@dot#1.#2@{#1}
\def\@set@refname#1{
    \edef\@tmp{\getrefbykeydefault{#1}{anchor}{}}%
    \xdef\@tmp{\expandafter\@throw@dot\@tmp.@}%
    \ltx@IfUndefined{\@tmp autorefnameplural}%
         {\def\@refname{\@nameuse{\@tmp autorefname}s}}%
         {\def\@refname{\@nameuse{\@tmp autorefnameplural}}}%
}
\def\@first@ref#1,#2{%
  \ifx#2@\autoref{#1}\let\@nextref\@gobble
  \else%
    (\ref{#1}
    \let\@nextref\@next@ref
  \fi%
  \@nextref#2%
}
\def\@next@ref#1,#2{%
   \ifx#2@,~\ref{#1}\let\@nextref\@gobble
   \else, \ref{#1}
   \fi%
   \@nextref#2%
   )
}
\makeatother

\usepackage{autonum}

\crefformat{equation}{(#2#1#3)}
\crefrangeformat{equation}{(#3#1#4) to~(#5#2#6)}

\AfterEndPreamble{%
\let\autoref\cref
}%

\makeatletter
\newcommand{\restore@Environment}[1]{%
  \AtBeginDocument{%
    \csletcs{#1*}{#1}%
    \csletcs{end#1*}{end#1}%
  }%
}
\forcsvlist\restore@Environment{alignat,equation,gather,multline,flalign,align}
\makeatother

  \renewcommand{\vec}[1]{\boldsymbol{#1}}
  \renewcommand{\v}[1]{\vec{#1}}
  \newcommand{\norm}[1]{\left\Vert#1\right\Vert}

  \newcommand{\paren}[1]{{\left(#1\right)}}
  \newcommand{\brc}[1]{\left\{#1\right\}}
  \newcommand{\brk}[1]{\left[#1\right]}
  
  \newcommand{\given}[2]{\left.#1\right\vert#2}

  \newcommand{\Cov}[2]{\text{Cov}\left(#1, #2\right)}

  \NewDocumentCommand\distn{mg}{
    #1\IfNoValueTF{#2}{}{\left(#2\right)}
  }

  \DeclareMathOperator{\GEV}{\distn{\text{GEV}}}
  \DeclareMathOperator{\IG}{\distn{\text{IG}}}
  \DeclareMathOperator{\GP}{\distn{\text{GP}}}

  \newcommand{\vs}{{\v s}}
  \newcommand{\vt}{{\v t}}
  \newcommand{\vx}{{\v x}}
  \newcommand{\D}{{\mathcal D}}
  \newcommand{\DS}{{\mathcal S}}

  \newcommand*{\nolink}[1]{%
    \begin{NoHyper}#1\end{NoHyper}%
  }

\title{Improved return level estimation via a weighted likelihood, latent spatial extremes model}

\usepackage{fancyhdr}
\usepackage{authblk}
\author{Joshua Hewitt}
\author{Miranda J. Fix}
\author{Jennifer A. Hoeting}
\author{Daniel S. Cooley}
\affil{Colorado State University}

\date{\vspace{-5ex}}

\IfFileExists{upquote.sty}{\usepackage{upquote}}{}
\begin{document}

\singlespace

\maketitle

\doublespace

\begin{abstract}
{\textbf{Abstract:}} Uncertainty in return level estimates for rare events,
like the intensity of large rainfall events, makes it difficult to develop
strategies to mitigate related hazards, like flooding. Latent spatial extremes
models reduce uncertainty by exploiting spatial dependence in statistical
characteristics of extreme events to borrow strength across locations. However,
these estimates can have poor properties due to model misspecification:
many latent spatial extremes models do not account for extremal dependence,
which is spatial dependence in the extreme events themselves. We improve
estimates from latent spatial extremes models that make conditional
independence assumptions by proposing a weighted likelihood that uses the
extremal coefficient to incorporate information about extremal dependence during
estimation. This approach differs from, and is simpler than, directly
modeling the spatial extremal dependence; for example, by fitting a max-stable
process, which is challenging to fit to real, large datasets.
We adopt a hierarchical
Bayesian framework for inference, use simulation to show the weighted
model provides improved estimates of high quantiles, and apply our model to
improve return level estimates for Colorado
rainfall events with 1\% annual exceedance probability.
\end{abstract}


\section{Introduction}
\label{sec:intro}

Natural hazards with potentially catastrophic impacts arise as extremes of
physical processes that are inherently dependent over space, such as large
storms that generate extreme precipitation. Accordingly, the statistical
modeling of spatially-referenced extreme values has been an active research
area in recent years.
To effectively plan mitigation strategies for natural hazards caused by extreme
precipitation, it is important to build maps that estimate
occurrence probabilities and return levels for extreme precipitation events at
individual locations. Return level maps for individual locations inform
building safety standards, insurance risks, and surface water runoff
requirements for stormwater management systems.
However, extreme events are rare by definition, so
relevant datasets from networks of environmental monitoring stations typically
have relatively short observation lengths. Spatial extremes models allow the
tails of probability distributions to be estimated while ``borrowing strength''
from neighboring time series.  Widely used to borrow strength, hierarchical
models share
statistical information across sampling locations to obtain more accurate
and spatially consistent estimates of extreme event characteristics.

Often in extremes studies, the primary interest is in modeling return levels of
extreme events at individual locations. Latent spatial extremes models are a
flexible and
computationally efficient class of models for marginal distributions of
spatial extremes and quantities derived from them, like return levels.
Latent spatial extremes models use a hierarchical framework to add spatial
structure to the {\it parameters} of an extreme value distribution.  Many
hierarchical frameworks assume observations of extremes are independent across
sampling locations, conditional on the latent spatial processes that specify the
data's marginal distributions. Hierarchical spatial layers induce smoothness
and correlation in marginal return level estimates across sampling locations,
and---critically---allow return level maps to be built
using spatial interpolation techniques, like kriging. As such, return level
estimates ``borrow strength'' because estimates balance data at each sampling
location with spatial smoothing induced by the latent hierarchical layers.
For example, \citet{Cooley2007a} use latent Gaussian processes in a
hierarchical Bayesian
model to capture covariate-driven trends and spatial dependence in precipitation
data. Bayesian frameworks allow direct estimation of uncertainties in return
levels since the posterior distribution contains this information.  Latent
spatial Gaussian process models can also be scaled to massive datasets with
recent advances in models and computational techniques
\citep{Lindgren2011a, Rue2009}.  Other recent studies employ latent spatial
extremes models in either Bayesian or frequentist paradigms
\citep{Sang2009, Cooley2010, Lehmann2016, Opitz2018a}. However, due to
the conditional independence assumption, these examples
of latent spatial extremes model cannot account for extremal dependence,
which is dependence in observations of extreme events themselves.

Directly modeling extremal dependence poses theoretical and computational
challenges. Classical univariate and multivariate extreme value models are
generated via asymptotic arguments about the limiting distributions of
appropriately renormalized block maxima. The natural extension to the spatial
setting is the max-stable process, which is the limiting process of the
componentwise maxima of a sequence of suitably renormalized stochastic
processes. Examples include the \cite{Smith1990}, \cite{Schlather2002}, and
Brown-Resnick \citep{Brown1977, Kabluchko2009} processes. The advantage of
max-stable process modeling is that it directly models spatial dependence in
the tail and thus permits inference about joint probabilities in addition to
marginal quantities, like return levels. However, full likelihood inference for
max-stable processes is only computationally tractable in relatively
low-dimensional situations \citep{Davison2012, Castruccio2016}.

In particular, computationally efficient Bayesian methods for
spatially-dependent extremes data remains challenging.
Frequentist inference for max-stable processes has typically been based on
computationally efficient models that use approximate likelihoods, such as
composite likelihoods based on bivariate densities of max-stable processes
\citep{Padoan2010}. However, composite likelihood methods are computationally
expensive and difficult to implement in hierarchical Bayesian models
\citep{Ribatet2012, Sharkey}.  Some Bayesian models do not need to use
approximate likelihoods, but are limited to specific max-stable processes or
require additional data for estimation \citep{Reich2012, Thibaud2016}.

The latent spatial extremes approaches previously introduced address
computational issues while providing flexible models for estimating marginal
parameters, but raise concerns about the impact of model
misspecification on inference.  These models make a simplifying conditional
independence assumption by defining the likelihood to be the product of each
location's marginal density. The misspecification due to the conditional
independence assumption can result in unrealistically narrow confidence
intervals for return level estimates \citep{Zheng2015, Cao2018}. Alternative to
assuming conditional independence or using computationally expensive models to
account for extremal dependence, we seek a compromise between the two modeling
approaches.  We want to preserve computationally efficient and flexible
models for marginal parameters provided by latent variable models, but
also account for extremal dependence in observations.

We propose a method for improving marginal inference that is supported by
theory and computationally efficient.  We develop a weighted likelihood
that uses spatial information to induce an effective sample size correction
that accounts for the loss of information due to dependent observations.
The likelihood weights improve uncertainty estimates in cases of
moderate to strong extremal dependence.  The effective sample size
motivation differs from previous uses of weighted likelihoods.  Weighted
likelihoods have previously been used to approximate Bayesian inference and as
a method for conducting inference on data sampled from multiple, related
populations, for example in \citet{Wang2006, Newton1994, Hu2002}.  Weighted
likelihoods have also recently been proposed for latent spatial extremes models,
but only as they relate to composite likelihood corrections \citep{Sharkey}.
A natural tradeoff in using likelihood weights to better account for estimation
uncertainty is that mean squared error can be slightly worse in these cases.

The remainder of the article is organized as follows. \Cref{sec:weighted}
introduces our weighted likelihood and Bayesian implementation.
\Cref{sec:simulation} uses a simulation study to show that the weighted
likelihood improves estimates, as compared to several models with similar
Bayesian hierarchical
structure.  As part of our comparisons, we derive the penalized complexity prior
for the generalized extreme value (GEV) distribution
(Supplement \nolink{\Cref{supp:sec:pc_derivation}}).
\Cref{sec:coflood} applies the weighted likelihood latent model to
daily rainfall observations in Colorado's Front Range of the Rocky Mountains.
We conclude with discussions of extensions and other directions for future
work (\Cref{sec:discussion}).

\section{Weighted likelihood latent spatial extremes models}
\label{sec:weighted}

We briefly review extreme value theory for modeling return levels from
observations of annual maxima (\Cref{sec:background}).  In particular, we
introduce the extremal coefficient, which we will use to build our weights.
We then propose and
interpret a latent variable model with a weighted likelihood to estimate
marginal quantities from spatially-dependent
extremes data (\Cref{sec:wtdlik}, \Cref{sec:ess_interp}).  When data are
dependent, the weighted
likelihood accounts for model misspecification in the latent variable
modeling approach by \citet{Cooley2007a}, which assumes data are conditionally
independent, given marginal parameters.  The model has a hierarchical spatial structure, for which posterior distributions can be
approximated via Gibbs sampling
(\Cref{sec:hierarchical}, \Cref{sec:estimation}).

\subsection{Max-stable processes and the extremal coefficient}
\label{sec:background}

Max-stable processes for spatially-referenced extremes data arise as the
pointwise limit of block maxima, which are pointwise maxima of
replications of spatially-referenced processes.  Let $\D$ be a continuous
spatial domain and $\brc{Y_{it}\paren\vs}_{\vs\in\D}$,
$t\in\brc{1,\dots,m}$ be $m$
independent replications of a spatial process at time block
$i\in\mathcal T = \brc{1,\dots,T}$.  The size of each block
$i\in\mathcal T$ is represented by $m$.  As the block size $m$ increases, if
the limit
\begin{align}
\label{eq:maxstable_lim}
    Y_i\paren\vs = \lim_{m\rightarrow\infty}
        \frac{\max_{t=1}^m Y_{it}\paren\vs - b_m\paren\vs}
             {a_m\paren\vs}, \vs\in\D
\end{align}
exists for continuous functions $a_m\paren\vs>0$ and $b_m\paren\vs\in\mathbb R$,
then $\brc{Y_i\paren\vs}_{\vs\in\D}$, $t\in\mathcal T$ are independent
replications of a max-stable process \citep{DeHaan1984}.

In general, the spatial dependence structure for
max-stable processes $\brc{Y_i\paren\vs}_{\vs\in\D}$ is complex, but is often
summarized for pairs of random variables $Y_i\paren\vs$ and $Y_i\paren\vt$
through the extremal coefficient.  The extremal coefficient $\theta\paren d$
is a function that is traditionally defined implicitly for
stationary and isotropic fields such that
\begin{align}
\label{eq:extcoeff}
    P\paren{Y_i\paren\vs \leq y, Y_i\paren\vt \leq y} =
    P\paren{Y_i\paren\vs \leq y}^{\theta\paren d}
\end{align}
for pairs of random variables $Y_i\paren\vs$ and $Y_i\paren\vt$ where
$d=\Vert\vs-\vt\Vert$ \citep{Schlather2003}.  The extremal coefficient is
interpretable as the effective number of independent random variables
among pairs of variables separated by a distance $d$.  As such, it takes values
in the closed interval $[1,2]$.

Importantly, the univariate marginal distributions for max-stable processes
belong to the generalized extreme value distribution family
$Y_i\paren\vs\thicksim\GEV\paren{\v\eta\paren\vs}$ with distribution function
\begin{align}
\label{eq:gev_cdf}
  P\paren{Y_i\paren\vs \leq y} = \begin{cases}
    \exp\brc{-\paren{1 +
      \xi\paren\vs\paren{\frac{y-\mu\paren\vs}{\sigma\paren\vs}}
    }^{-1/\xi\paren\vs}_+} & \xi\paren\vs \neq 0
    \\
    \exp\brc{-\exp\brc{\frac{y-\mu\paren\vs}{\sigma\paren\vs}} } &
        \xi\paren\vs = 0
  \end{cases}
\end{align}
where $a_+ = \max\paren{0, a}$ \citep{DeHaan1984}. The parameter vector
$\v\eta\paren\vs = \paren{\mu\paren\vs, \log\sigma\paren\vs, \xi\paren\vs}^T$
specifies the distribution's location $\mu\paren\vs \in\mathbb R$, scale
$\sigma\paren\vs > 0$, and shape $\xi\paren\vs \in\mathbb R$ parameters.
The GEV quantile function $Q\paren{\given{p}{\v\eta\paren\vs}}$ is derived from
\autoref{eq:gev_cdf} and has the closed form
\begin{align}
\label{eq:gevquantile}
  Q\paren{\given{p}{\v\eta\paren\vs}} = \begin{cases}
    \mu\paren\vs +
      \frac{\sigma\paren\vs}{\xi\paren\vs}
      \paren{\paren{-\log p}^{-\xi\paren\vs} - 1} & \xi\paren\vs\neq 0 \\
    \mu\paren\vs - \sigma\paren\vs\log\paren{-\log p} & \xi\paren\vs = 0
  \end{cases}
\end{align}
with $p\in\brk{0,1}$.

Asymptotic
convergence justifies use of the GEV distribution as an approximate model for
the annual maximum of daily precipitation in year $i$, which is a block maximum
quantity that has large but finite replication $t\in\brc{1,\dots,m}$.  The
approximation allows marginal return levels for extreme precipitation events to
be modeled as high quantiles of the GEV distribution at each location
$\vs\in\D$. Assuming a stationary climate, the quantile
$Q\paren{\given{p}{\v\eta\paren\vs}}$ with $p=1-1/r$ is interpretable as the
$r$-year return level---the amount of precipitation carried by a storm that
occurs, on average, once every $r$ years.  The quantile
$Q\paren{\given{p}{\v\eta\paren\vs}}$ is also associated with the $1-p$ percent
annual exceedance probability; the quantile expresses the amount of
precipitation carried by a storm that has a $1-p$ percent chance of occurring
in a given year.

\subsection{Weighted likelihood}
\label{sec:wtdlik}

We propose a latent variable model that uses a weighted marginal likelihood.
In general, weighted likelihoods are missspecified but can improve inference
relative to unweighted likelihoods. A correctly-specified likelihood for
spatial extremes data would fully account for extremal dependence, but be
computationally intractable.
Marginal likelihoods assume data are conditionally independent across spatial
locations and timepoints, given marginal parameters. When the field
$\brc{Y_i\paren\vs}_{\vs\in\D}$ is sampled at
$N$ spatial locations $\DS = \brc{\vs_1, \dots, \vs_N} \subset \D$, the
weighted marginal likelihood for a finite sample of observations
$\brc{y_i\paren{\vs_j} : i\in\mathcal T, \vs_j\in\DS}$ is defined via
\begin{align}
\label{eq:wtdLik}
  L\paren{\v\eta} = \prod_{j=1}^N \prod_{i=1}^T
    f\paren{\given{y_i\paren{\vs_j}}{\v\eta\paren{\vs_j}}}^{w_{\vs_j}}
\end{align}
where $f\paren{\given{y_i\paren{\vs_j}}{\v\eta\paren{\vs_j}}}$ is the
probability density function for the GEV distribution \autoref{eq:gev_cdf}
and $\v\eta\paren\vs$ is the associated parameter vector.
The weighted marginal likelihood \autoref{eq:wtdLik} uses likelihood weights
$\brc{w_{\vs_j}:j=1,\dots,N}$ and marginal densities
$\brc{f\paren{\given{y_i\paren{\vs_j}}{\v\eta\paren{\vs_j}}} : j=1,\dots,N}$
to estimate the marginal parameters
$\brc{\v\eta\paren{\vs_j} \in\mathbb R^3 : j=1,\dots, N}$ that have been
stacked to form the vector $\v\eta\in\mathbb R^{3N}$.  During estimation,
likelihood weights can be constructed to downweight observations for
$y_i\paren{\vs_j}$ that exhibit strong dependence with neighboring observations.
Models assuming conditional independence naively assume the weights are unitary.

We use the extremal coefficient in \autoref{eq:extcoeff} to construct weights
that downweight likelihood contributions from locations central to
the spatial sampling pattern, where observations tend to be most dependent.  We
construct each weight $w_{\vs_j}$ by first mapping extremal coefficients
$\theta\paren{\Vert\vs_i -\vs_j\Vert}$ for $i\neq j$ to the interval
$\brk{1/N,1}$, then averaging the mapped values, yielding
\begin{align}
\label{eq:wts}
  w_{\vs_j} = \frac{1}{N-1} \sum_{i=1, ~i\neq j}^N
    N^{\theta\paren{\Vert\vs_i - \vs_j\Vert} - 2},
\end{align}
so $w_{\vs_j}\in\brk{1/N,1}$.
The weights
\autoref{eq:wts} are specifically constructed so that the statistical
information in the weighted marginal likelihood \autoref{eq:wtdLik} matches the
statistical information in non-misspecified likelihoods
in two special, limiting cases
(Supplement \nolink{\Cref{supp:sec:wts_motivation}}).
In the first limiting case, the field is assumed to be independent, and
$w_{\vs_j}=1$; in the second limiting case,  the field
$\brc{Y_i\paren\vs}_{\vs\in\D}$ is assumed to have complete dependence over
space, and $w_{\vs_j}=1/N$. The field $\brc{Y_i\paren\vs}_{\vs\in\D}$ has
complete dependence over space if all potential samples
$\brc{Y_i\paren{\vs_1},\dots,Y_i\paren{\vs_N}}$ can be represented through a
collection of continuous transformations $\brc{g_j:j=1,\dots,N}$ of a variable
$U_i$ such that
\begin{align}
    \paren{Y_i\paren{\vs_1},\dots,Y_i\paren{\vs_N}} \overset{d}{=}
      \paren{g_1\paren{U_i},\dots, g_N\paren{U_i}}.
\end{align}

\subsection{Effective sample size interpretation}
\label{sec:ess_interp}

From an information-theoretic perspective, we show that the weighted likelihood
\autoref{eq:wtdLik} induces an effective sample size that corrects inference on
spatially correlated marginal parameters when data are also spatially dependent.
Effective sample size statistics quantify the impact that dependence has on
estimation uncertainty \citep[e.g.,][p. 13]{Cressie1993}.  We use effective
sample size to determine factors that will impact estimator performance in our
simulation (\Cref{sec:simulation}).  In our application, effective sample size
also helps us better interpret losses in statistical efficiency due to
dependence in observations of extremes (\Cref{sec:coflood}).

The Fisher information for \autoref{eq:wtdLik} is the block diagonal matrix
$I\paren{\v\eta}\in\mathbb R^{Nm\times Nm}$ with $j^{th}$
diagonal block $I\paren{\v\eta\paren{\vs_j}} \in\mathbb R^{m\times m}$ being
\begin{align}
\label{eq:wtdFI}
    I\paren{\v\eta\paren{\vs_j}} =
        w_{\vs_j} T I_{Y_\bullet\paren{\vs_j}}\paren{\v\eta\paren{\vs_j}},
\end{align}
where $I_{Y_\bullet\paren{\vs_j}}\paren{\v\eta\paren{\vs_j}}$ is the expected
Fisher information for each of the independent and identically distributed
random variables $\brc{Y_i\paren{\vs_j} : i\in\mathcal T}$.
Note that the $j^{th}$ block \autoref{eq:wtdFI} is the Fisher information for
$w_{\vs_j} T$ independent observations of the response at $\vs_j$.
Thus, $w_{\vs_j}$ quantifies the effective proportion of
independent observations at location $\vs_j$ that contribute to inference for
the marginal GEV parameters $\v\eta\paren{\vs_j}$.  As the likelihood weight
$w_{\vs_j}$ decreases, uncertainty increases about the marginal GEV parameters
$\v\eta\paren{\vs_j}$ and return level
$Q\paren{\given{p}{\v\eta\paren{\vs_j}}}$.  Latent spatial extremes models with
unweighted likelihoods can be interpreted as implicitly assigning
$w_{\vs_j}=1$ for all locations $\vs_j\in\DS$.  Such a strategy
underestimates parameter uncertainty when data have extremal dependence.

\subsection{Hierarchical specification}
\label{sec:hierarchical}

We adopt a hierarchical Bayesian framework to conduct inference on the weighted
marginal likelihood, and facilitate spatial interpolation of marginal return
levels \autoref{eq:gevquantile}.  We specify a hierarchical spatial process
model for the marginal parameters at each spatial location \linebreak
$\v\eta\paren\vs = \paren{\mu\paren\vs, \log\sigma\paren\vs, \xi\paren\vs}^T
    \in\mathbb R^3$
via
\begin{align}
\label{eq:mean_form}
    \v\eta\paren\vs =
        \begin{bmatrix}
            \vx_\mu\paren\vs^T &  &  \\
             & \vx_{\log\sigma}\paren\vs^T &  \\
             &  & \vx_\xi\paren\vs^T
        \end{bmatrix}
        \begin{bmatrix}
            \v\beta_\mu \\ \v\beta_{\log\sigma} \\ \v\beta_\xi
        \end{bmatrix}
        +
        \begin{bmatrix}
            \varepsilon_\mu\paren\vs \\
            \varepsilon_{\log\sigma}\paren\vs \\
            \varepsilon_\xi\paren\vs \\
        \end{bmatrix},
\end{align}
in which $\vx\paren\vs$ and $\v\beta$ are respectively $p\times 1$ vectors of
regression covariates
and coefficients, and $\varepsilon\paren\vs$ represents spatially-correlated
variation in the marginal parameters $\v\eta\paren\vs$.  The matrix of
covariates in \autoref{eq:mean_form} is block-diagonal; the blank, off-diagonal
entries represent zeros.
We use diffuse normal priors for regression coefficients $\v\beta$.
Independent Gaussian processes model
the spatially-correlated variation in $\varepsilon_\mu\paren\vs$,
$\varepsilon_{\log\sigma}\paren\vs$, and $\varepsilon_\xi\paren\vs$.
Gaussian processes imply finite samples of parameters are jointly-normally
distributed and allow estimation of spatially-coherent marginal parameter
maps $\brc{\v\eta\paren{\vs}}_{\vs\in\D}$ through kriging. Furthermore,
stationary isotropic Gaussian processes are sufficient models when departures
from stationarity and isotropy are difficult to detect \citep{Cooley2007a}.

The Gaussian processes for marginal parameters are fully defined by specifying
covariance functions
$\Cov{\varepsilon\paren\vs}{\given{\varepsilon\paren{\v t}}{\v\phi}} =
    \rho\paren{\norm{\vs-\v t}; \v\phi}$
to model the spatial correlation in the parameters between locations
$\vs,\v t\in\D$.  Specific choices for covariance functions $\rho$ and
hyperprior distributions for covariance parameters
$\v\phi = \paren{\sigma_0, \lambda_0, \nu_0}^T$ are discussed in
\Cref{sec:sim_model} and \Cref{sec:co_model}.  In general, we use weakly
informative Gamma priors for covariance range $\lambda_0$ and smoothness
$\nu_0$ parameters, and weakly informative Inverse-Gamma priors for covariance
sill parameters $\sigma_0$.

\subsection{Bayesian estimation}
\label{sec:estimation}

A Gibbs sampler can be constructed for inference on the hierarchical Bayesian
model specified in \Cref{sec:hierarchical}, in which likelihood weights
\autoref{eq:wts} are updated with the aid of a plug-in estimator for the
extremal coefficient.  The Bayesian framework allows
estimates of return levels $Q\paren{\given{p}{\v\eta\paren\vs}}$ to be computed
directly from posterior samples of the marginal parameter vector
$\v\eta\paren\vs$ since return levels are functions of marginal parameters.
The sampler is described in detail in Supplement
\nolink{\Cref{supp:sec:gibbs}}, and key points are summarized here.  Standard
hybrid Gibbs sampling approaches are used to sample the marginal GEV parameters,
covariance parameters, and regression coefficients.

Likelihood weights \autoref{eq:wts} are computed with a plug-in estimator
$\hat\theta\paren d$ for the extremal coefficient \citep{Cooley2006}.  The
plug-in estimator uses sample statistics from the data that have
been transformed to have unit Fr\'{e}chet marginal distributions.  Thus, the
likelihood weights depend on estimates of the marginal distributions, either
estimated through the empirical cumulative distribution function (CDF), or
directly through the GEV CDF.  Before Gibbs sampling begins, we initialize
likelihood weights by using the empirical CDF at each location
$\hat F\paren{y; \vs_j} = T^{-1} \sum_{i=1}^T
    \mathds{1}\brc{y_i\paren{\vs_j} \leq y} $
to transform the data via probability integral transforms.
These initial weights may be held fixed and used throughout
Gibbs sampling or updated at each Gibbs iteration.  To update the weights at
each Gibbs iteration, the data may be retransformed by using the GEV CDF
\autoref{eq:gev_cdf} with the marginal parameters $\v\eta$ from the previous
Gibbs iteration. Updating likelihood weights during Gibbs sampling accounts for
uncertainty in the likelihood weights.

\section{Simulation study}
\label{sec:simulation}

We use simulation to show that the weighted marginal likelihood
\autoref{eq:wtdLik} improves high quantile estimates on datasets with realistic
GEV parameters $\v\eta\paren\vs$, sample sizes, and varying extremal
dependence. The simulation compares the weighted likelihood model
(\Cref{sec:comparison_models}) to a standard, unweighted latent spatial
extremes model and a penalized variation.  Penalization is an
alternate approach used to correct return level estimates in extreme value
models \citep[cf.][]{Schliep2010, Opitz2018a}.  Penalized models have
hierarchical structures that are similar to our weighted likelihood, so are
comparison models with similar computational complexity to our weighted
likelihood.  We compare models by
contrasting properties of estimators of high quantiles, including empirical
coverage and mean squared error (\Cref{sec:simulation_results}).

\subsection{Datasets}
\label{sec:simulation_data}

We simulate data from four generating models with varying combinations of
extremal dependence, and spatial $N$ and temporal $T$ sample sizes.  Properties
of parameter estimators are empirically approximated using 1,000 datasets
simulated from each generating model. Our decision to vary extremal dependence,
$N$, and $T$ is informed by the Fisher information \autoref{eq:wtdFI} and
effective sample size discussion (\Cref{sec:wtdlik}), which provide intuition
about how extremal dependence and sample size affect estimation.
Increasing extremal dependence decreases the amount of statistical information
available for parameter estimation, much as occurs with classical spatial
dependence \citep[Section 1.3]{Cressie1993}.  Similarly, the impact of extremal
dependence increases when sampling more spatial locations
$\mathcal S = \brc{\vs_1,\dots,\vs_N} \subset\D$ from a fixed domain $\D$.
Unweighted latent spatial extremes models are misspecified when data are
dependent because they assume the data are conditionally independent given
model parameters.
The severity of the misspecification increases as the process is observed at
more spatial locations $N$ because it becomes more likely that observations
from spatially-dependent locations are included in the sample.  The Fisher
information equation \autoref{eq:wtdFI}, however, suggests that statistical
information about the marginal parameters increases with the number of
replications $T$ despite misspecification, albeit at a slower rate when
using likelihood weights.

Simulated data have marginal GEV parameters $\v\eta\paren\vs$ that mimic
estimates from observed annual maximum daily precipitation across Colorado's
Front Range \citep{Tye2015}.  Spatially-dependent GEV parameters
$\v\eta\paren\vs$, $\vs\in\D=\brk{-10,10}^2$ are sampled from Gaussian processes
$\GP\paren{m, \, \rho}$  with mean functions $m:\D\rightarrow\mathbb R$ and
powered exponential covariances $\rho:\D^2\rightarrow\left[0,\infty\right)$
specified in \Cref{table:sim_params}.  Shape parameters $\xi\paren\vs$
are resampled until $\xi\paren\vs>0$ for all $\vs\in\DS$ to ensure data are
heavy-tailed. Brown--Resnick processes model extremal dependence in the
simulated data \citep{Kabluchko2009}.  The semi-variogram
$\gamma : \D^2 \rightarrow \left[0,\infty\right)$ specified in
\Cref{table:sim_params} parameterizes a Brown--Resnick model that induces
strong, medium, or weak extremal dependence on $\D$ as measured by
the extremal coefficient function $\theta\paren d$.  For comparison,
independent data are also simulated.

\begin{table*}\centering
\caption{Generating model configurations used to simulate data for comparing
    the weighted likelihood \autoref{eq:wtdLik} to alternate estimating models
    (\Cref{sec:comparison_models}).  We evaluate model performance with 1,000
    datasets for each combination of spatial $N$ and temporal $T$ sample sizes,
    and extremal dependence.}
\label{table:sim_params}
\ra{2}
\begin{tabularx}{\linewidth}{@{}>{\em}r@{\hspace{1.5em}}X@{}} \toprule

Spatial sample size & \makecell[l]{$N \in \brc{ 30, \, 50, \, 100}$ sites
  sampled uniformly on $\D = \brk{-10, \, 10}^2$} \vspace{-1em} \\

Temporal sample size & $T \in\brc{50, \, 100}$ \vspace{.5em} \\

\makecell[tr]{Extremal dependence \\ {\footnotesize (Brown-Resnick parameters)}} &
\ra{1.125}
\begin{tabular}[t]{@{}ll}
    Semi-variogram & $\gamma_\paren{\lambda, ~\alpha}\paren{\vs_1, \vs_2} =
      \paren{\Vert\vs_1-\vs_2\Vert/\lambda}^\alpha$ \\
      Independent & $\paren{\lambda=\text{NA}, ~\alpha=\text{NA}}$ \\
      Weak & $\paren{\lambda=.25, ~\alpha=.75}$ \\
      Moderate & $\paren{\lambda=.5, ~\alpha=.5}$ \\
      Strong & $\paren{\lambda=.75, ~\alpha=.25}$ \\
\end{tabular}
\ra{2} \vspace{1em} \\

\makecell[r]{Prior distributions \\
  for GEV parameters $\v\eta\paren\vs$} &
\begin{itemize}[leftmargin=0pt, label={}, itemsep=-.3em]
  \vspace{-2.5em}
  \item \raggedright Covariance function \linebreak
    $\rho_{\paren{\sigma_0, ~\lambda_0, ~\nu_0}}\paren{\vs_1, \vs_2} =
        \sigma_0 \exp\brc{-\paren{\Vert\vs_1 - \vs_2\Vert/\lambda_0}^{\nu_0}}$
     \vspace{.25em}
  \item \raggedright Gaussian processes \linebreak
   $\mu\paren{\vs} \thicksim \GP
    \paren{26 + \brk{.5 ~~ 0}^T \vs,
        ~ \rho_\paren{4, ~20, ~1}}$
  \item $\log\sigma\paren{\vs} \thicksim \GP
    \paren{\log(10) + \brk{0 ~~ .05}^T \vs, ~ \rho_\paren{.4, ~5, ~1}}$
  \item $\xi\paren{\vs} \thicksim \GP
    \paren{.12, ~ \rho_\paren{.0012, ~10, ~1}}$
\vspace{-.75em} \end{itemize}  \\

\bottomrule
\end{tabularx}
\end{table*}

\subsection{Estimating models}
\label{sec:comparison_models}

The simulation compares estimation of conditionally independent models with
weighted \autoref{eq:wtdLik} and unweighted likelihoods (i.e.,
\autoref{eq:wtdLik} with $w_{\vs_j}=1$ for all $\vs_j\in\DS$) and a variation
that uses penalized complexity priors as a likelihood penalty
(\Cref{sec:pc_prior}).  Key
differences between the estimating models are summarized in
\Cref{table:comp_models}. The comparison models represent different approaches
proposed in the extremes literature to improve marginal estimation of GEV
parameters and have similar computational complexity.

\subsubsection{Penalized complexity prior}
\label{sec:pc_prior}

Likelihood-based parameter estimates for the univariate GEV distribution are
known to perform poorly, but penalized likelihoods can
reduce estimation bias \citep{Coles1999, Martins2000}.  Penalized likelihoods
have been incorporated into spatial models for marginal extremes
\citep{Schliep2010, Opitz2018a}.  Penalization improves estimation of
marginal parameters by downweighting estimates of large shape parameters
$\xi\paren\vs$, which tend to be uncommon in many extreme precipitation data.
We adapt a contemporary penalty for use with the GEV distribution as a
comparison model.

Penalized complexity (PC) priors have recently been proposed to improve
parameter estimation in a related extreme value family---the Generalized
Pareto distribution (GPD), which also uses scale $\sigma\paren\vs>0$, and
shape $\xi\paren\vs\in\mathbb R$ parameters to model threshold exceedances
\citep{Opitz2018a}. Penalized
complexity priors satisfy several properties that optimize the prior
distribution's shape and scale to precisely control the prior's influence over
target likelihoods \citep{Simpson2017}. We implement PC priors as
penalized likelihoods in our hierarchical spatial model.  We derive the
penalized complexity prior $\pi\paren{\given{\xi}{\lambda}}$ for the GEV
distribution in Supplement \nolink{\Cref{supp:sec:pc_derivation}} and use it
with the log-likelihood
\begin{align}
\label{eq:pclik}
    \ell\paren{\v\eta} = \sum_{j=1}^N \sum_{i=1}^T
        \log f\paren{\given{y_i\paren{\vs_j}}{\v\eta\paren{\vs_j}}} +
        \sum_{j=1}^N \log \pi\paren{\given{\xi\paren{\vs_j}}{\lambda}}
\end{align}
in place of the log of the unweighted version of the likelihood
\autoref{eq:wtdLik}, in which $w_{\vs_j}=1$ for all $\vs_j\in\DS$.

Bayesian estimation optimizes the PC prior's parameterization by specifying an
Inverse-gamma prior distribution for $\lambda \thicksim \IG\paren{2,1}$.
The Inverse-gamma distribution is parameterized to have mean 1 and infinite
variance.  Prior distributions provide an alternative to cross-validation
approaches for optimizing the prior's parameterization, which is
computationally infeasible for this simulation study \citep{Park2008, Hans2009}.

\begin{table*}\centering
\caption{Summary of differences between estimating models in simulation study
    (\Cref{sec:simulation}).}
\label{table:comp_models}
\ra{1.4}
\begin{tabularx}{\linewidth}{YccY} \toprule

{\em Model} &
{\em (Log-)Likelihood} &
{\em Weights} &
{\em Log-Likelihood penalty} \\

Unweighted & ~\autoref{eq:wtdLik} & None & None \\

Weighted & ~\autoref{eq:wtdLik} & \autoref{eq:wts} & None \\

PC Prior & ~\autoref{eq:pclik} & None &
    $\sum_j \log \pi\paren{\given{\xi\paren{\vs_j}}{\lambda}}$ \\

\bottomrule
\end{tabularx}
\end{table*}

\subsubsection{Bayesian specification}
\label{sec:sim_model}

All models use
a hierarchical Bayesian framework in which the GEV parameters $\v\eta\paren\vs$
are estimated as independent latent Gaussian processes with functional forms
matching those specified in \Cref{table:sim_params}.  Prior distributions for
the mean and covariance function parameters are either weakly informative or
uninformative, and conjugate where possible.  Full details are available in
Supplement \nolink{\Cref{supp:sec:priors}}.
Inference is based on a sample from the posterior distribution, drawn with a
Gibbs sampler.  Estimators based on the weighted likelihood are evaluated with
respect to both fixed and Gibbs-updated weights (See \Cref{sec:estimation}).
Sample autocorrelation
diagnostics indicate the Gibbs sampler mixes slowly, so the sampler was run
for 155,000 iterations to ensure Monte Carlo integration error is sufficiently
small.  The first 5,000 samples were discarded.  Posterior inference uses a
thinned posterior sample consisting of 10,000 of the remaining 150,000 samples;
only every fifteenth sample was saved because the entire posterior sample could
not be efficiently stored and manipulated.  Thinning reduces statistical
efficiency of Markov chain Monte Carlo methods, but can be a necessary tradeoff
when the full posterior sample is difficult to store and use to estimate
posterior quantities \citep{MacEachern1994}.

\subsection{Results}
\label{sec:simulation_results}

Assuming a stationary climate, the 1\% annual exceedance probability
$Q\paren{\given{.99}{\v\eta\paren\vs}}$ from \autoref{eq:gevquantile},
also referred to as the 100-year return level, is often used to quantify
risk for extreme weather events.  The weighted model's results
are nearly identical when comparing fixed weights to Gibbs-updated weights, so
we only present the fixed-weight results here; results for the Gibbs-updated
weights are included in Supplement \nolink{\Cref{supp:sec:sim_results}}.
\Cref{fig:rl_coverage} presents the empirical coverage of highest posterior
density (HPD) intervals for the return level
$Q\paren{\given{.99}{\v\eta\paren\vs}}$ for each of the models listed in
\Cref{table:comp_models}.  Supplement \nolink{\Cref{supp:fig:rl_mse}} presents
mean squared error (MSE) for the same data.
Bias is small for all estimators, so MSE mainly quantifies estimator variance.
Since the return level $Q\paren{\given{.99}{\v\eta\paren\vs}}$ is greatly
influenced by the shape parameter, $\xi\paren\vs$, results for return levels
and shape parameters are very similar. Supplement
\nolink{\Cref{supp:sec:sim_results}} includes results for all GEV parameters
$\v\eta\paren\vs$ and other estimator properties.

Extremal dependence degrades the performance of all marginal models, but the
weighted marginal likelihood \autoref{eq:wtdLik} provides the most accurate
estimates of uncertainty. Empirical coverage of 95\% HPD intervals is closest
to the nominal HPD level across all levels of extremal dependence.
(\Cref{fig:rl_coverage}).  For the $N=50, T=50$ simulation with moderate
dependence, the weighted model has a coverage rate of 86\%, while the
unweighted model and penalized complexity prior model have coverage rates of
83\% and 82\% respectively. In the same scenario, the weighted model also has
nearly identical MSE as the other models, although the MSE for the weighted
likelihood model is somewhat greater for the simulation with strong dependence
(Supplement \nolink{\Cref{supp:fig:rl_mse}}).

\begin{knitrout}
\definecolor{shadecolor}{rgb}{0.969, 0.969, 0.969}\color{fgcolor}\begin{figure}

{\centering \includegraphics[width=\maxwidth]{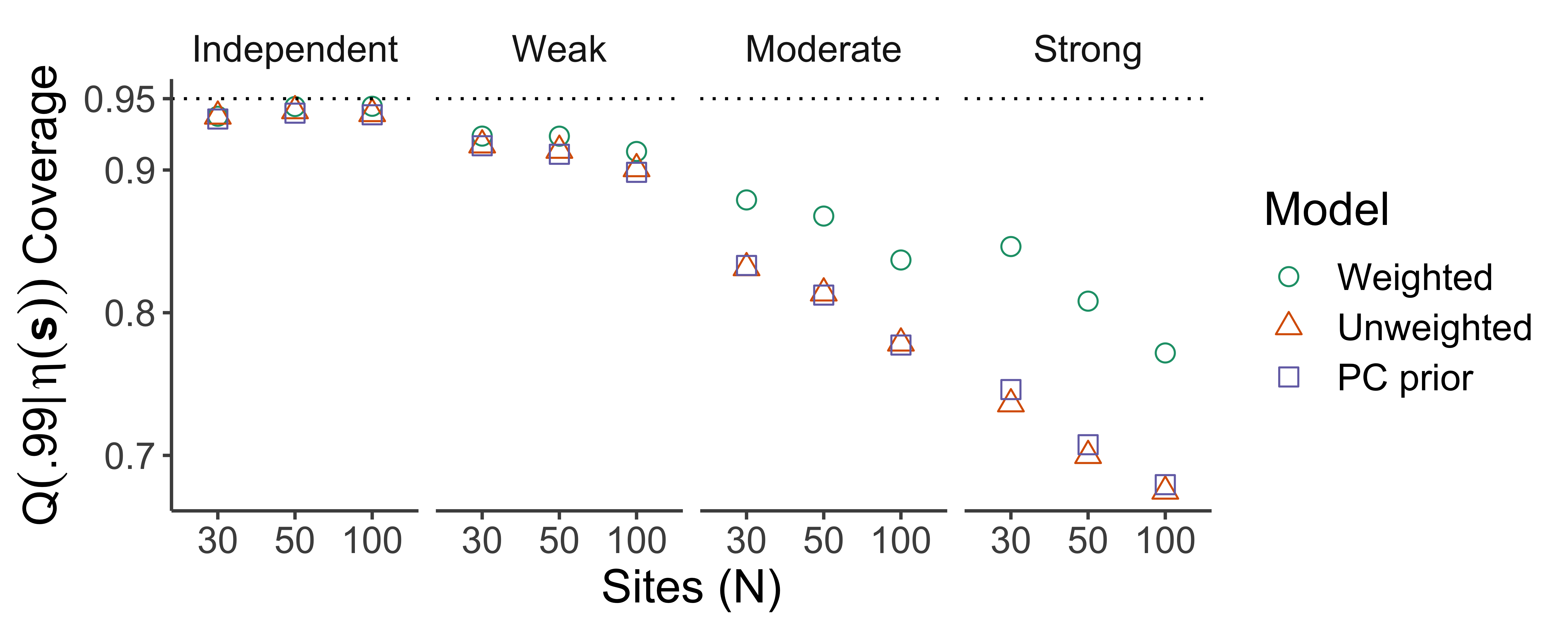} 

}

\caption{Empirical coverage rates of 95\% highest posterior density intervals for 100-year return levels $Q\paren{\given{.99}{\v\eta\paren{\vs}}}$ for four levels of extreme dependence across comparison models and simulations with $T=50$ observations per location. Nominal coverage is marked by the dotted horizontal reference line at .95.  While empirical coverage degrades for all estimating models as extremal dependence increases, the weighted model is most robust to model misspecification caused by extremal dependence.  Supplement \Cref{supp:sec:sim_results} includes results for $T=100$, which show slight improvement in all coverage rates.}\label{fig:rl_coverage}
\end{figure}

\end{knitrout}

\section{Extreme Colorado precipitation}
\label{sec:coflood}

\subsection{Data}

Previous studies of extreme precipitation in Colorado find that there is weak
extremal dependence between locations along the
state's Front Range region \citep{Cooley2007a, Tye2015}.  We determine the
impact the weighted likelihood \autoref{eq:wtdLik} has on estimates of the
1\% annual exceedance probability
$Q\paren{\given{.99}{\v\eta\paren\vs}}$, also referred to as the  100-year
return level. Estimates are based on the same subset of annual maxima of
daily precipitation \citet{Tye2015} use from the Global Historical Climatology
Network (GHCN) dataset \citep{Menne2012}.  The subset includes annual maxima
from 71 stations along the Front Range.  \citet{Tye2015} fully describe their
selection criteria, which, for example, include requirements that stations have
been operational for at least 30 years. Additionally, annual maxima of daily
precipitation are only analyzed from years with few missing daily records of
precipitation.  Between 18 and 120 annual maxima are analyzed for each
station, with roughly equal representation of all temporal sample sizes.

Exploratory analysis suggests the Front Range GHCN data have between weak and
moderate extremal dependence. The estimated extremal coefficient function
$\hat\theta\paren d : \paren{0,\infty} \rightarrow \brk{1,2}$ is near-constant
between 1.8 and 1.9 for all distances $d$, which implies the likelihood weights
will also have a small range.  \citet{Schlather2003} also observe a
near-constant extremal coefficient function for extreme precipitation in
south-west England.  The authors remark that the result may have a physical
basis because the study region is small relative to the scale of the
meteorological systems that generate precipitation, which implies it is likely
that no two sites in the region are truly independent.  Likelihood weights
\autoref{eq:wts} for the GHCN data are similar to weights for simulated
data with moderate extremal dependence
(Supplement \nolink{\Cref{supp:fig:co_wtd_need}}).
Since the average number of annual maxima per station ($T=60$) is also close to
our $T=50$ simulation, we anticipate the weighted likelihood will have
closer to nominal coverage and the unweighted likelihood will slightly
undercover (\Cref{fig:rl_coverage}).

\subsection{Model and results}
\label{sec:co_model}

As in the simulation, we use the weighted marginal likelihood
\autoref{eq:wtdLik} in a hierarchical Bayesian framework in which the GEV
parameters $\v\eta\paren\vs$ are estimated as independent latent Gaussian
processes.  Since the simulation shows that estimators based on fixed and
Gibbs-updated weights have similar properties, we use fixed weights during
estimation.  We use annual mean precipitation from the PRISM precipitation
dataset \citep{Daly2008} as a covariate for each of the GEV parameters, and
model the spatial correlation between parameters with the Mat\'{e}rn covariance
function.  For example, the Mat\'{e}rn specifies the correlation between
parameters $\xi\paren\vs$ and $\xi\paren\vt$ at two locations $\vs,\vt\in\D$ via
\begin{align}
    \kappa\paren{\vs,\vt;\tau, \rho, \nu} =
        \frac{1}{\tau2^{\nu-1}\Gamma\paren{\nu}}
        K_\nu\paren{\Vert\vs-\vt\Vert/\rho}
\end{align}
where $K_\nu$ is the modified Bessel function of the second kind with order
$\nu$.  The Mat\'{e}rn covariance is parameterized through its inverse scale
$\tau>0$, range $\rho>0$, and smoothness $\nu>0$ parameters. Annual
average precipitation from the PRISM dataset accounts for average weather
patterns and orographic effects on precipitation, such as elevation.
Prior distributions for the mean and covariance function parameters are
available in Supplement \nolink{\Cref{supp:sec:co_priors}}.  In general,
prior distributions are weakly informative, and prior distributions for spatial
covariance parameters are centered around variogram-based estimates of spatial
correlation between exploratory estimates of marginal parameters
$\v\eta\paren\vs$.

Inference uses a sample from the posterior distribution, drawn with a Gibbs
sampler that was run for 3,002,000 iterations.  The first 2,000 samples were
discarded. The sampler was run for a large number of iterations because it was
slowly mixing.  Posterior inference uses 10,000 of the remaining samples; only
every $300^{th}$ sample was saved due to storage constraints.
To facilitate model comparison, we also fit the unweighted latent
spatial extremes model using the same priors and inference strategy.
Posterior diagnostics for the weighted likelihood model
are presented in \Cref{supp:sec:co_diagnostics}.
Diagnostics suggest no significant concerns with convergence and also that the
chain has been run for long enough to control Monte Carlo integration error.
Due to the relatively
small number of spatial locations in the dataset ($N=71$), posterior
diagnostics indicate the spatial covariance parameters are at least weakly
identified by the data.  Posterior learning is diagnosed by comparing prior
and posterior distributions for the spatial mean and covariance parameters.

The likelihood weights \autoref{eq:wts} have a spatial pattern and their
effect can be interpreted by their impact on the weighted Fisher information
\autoref{eq:wtdFI} (\Cref{fig:co_wts_deltas}).  As expected, stations near the
edges of the sampled region
tend to have the highest weights because annual maxima observed at these
locations are at most weakly dependent with observations at other stations.
 Annual maxima at distant stations tend to be at
most weakly dependent because they tend to experience different large rain
events than other stations.

Weighted estimates borrow more strength across locations, which impacts return
level estimates.  The latent Gaussian processes increase
smoothing as more strength is borrowed, shrinking parameter estimates
(Supplement \nolink{\Cref{supp:fig:co_shapes_and_equivs}}).  Shrinkage
manifests as additional smoothing in maps of return levels
(\Cref{fig:co_spatial_diffs}).  In particular, the weighted estimates better
match physical features that impact Colorado precipitation.  The contours in
the weighted return level map have stronger north-south patterns, especially
along $105^\circ$ W---the boundary of the Rocky mountains in the Colorado Front
Range region (\Cref{fig:co_spatial_diffs} B).  The size of the region with
elliptical 150--175mm return level contours ($\color{contourblue}\blacksquare$)
of extreme precipitation near Boulder, Fort Collins, and Colorado Springs also
increase. The larger elliptical regions produced by the weighted model better
capture physical effects of the Palmer Divide and the Cheyenne Ridge on
Colorado precipitation \citep{Daly2008, Karr1976}.

We verify that the weighted model's changes are beneficial near the Palmer
Divide and Cheyenne Ridge regions by refitting the weighted and unweighted
models with a holdout set to
test out-of-sample fit.  Our holdout set uses data from seven stations
(10\% of the dataset) near the Palmer Divide and Cheyenne Ridge, and where
posterior estimates of return levels differ between the two models (stations
marked by diamonds in \Cref{fig:co_wts_deltas}).  Testing uses the log-score
$\ell\paren{\vs_0}$ at each holdout location $\vs_0$.  Log-scores form
strictly proper scoring rules that compare the log-likelihood from both models
on data at each holdout location \citep{Gneiting2007a}.  In our spatial
application, we use the posterior kriging distribution to draw a posterior
sample of GEV parameters at each test location $\vs_0$, which we then use to
compute the posterior mean log-likelihood at each test location
$\ell\paren{\vs_0}$.  Resulting log-scores show that the weighted model
improves out-of-sample fit in six out of seven of the holdout locations
(\Cref{table:co_holdout_table}).  The log-scores also show that neither model
fits the data well at Pueblo, CO, the southernmost holdout station.  In
particular, the data at Pueblo, CO tend to be relatively less extreme.
Separate exploratory analysis of Pueblo's data suggests extreme precipitation
is associated with a negative shape parameter $\xi\paren{\vs_0}<0$.  However,
the spatial models suggest a positive shape parameter is more appropriate.

\begin{knitrout}
\definecolor{shadecolor}{rgb}{0.969, 0.969, 0.969}\color{fgcolor}\begin{figure}

{\centering \includegraphics[width=\maxwidth]{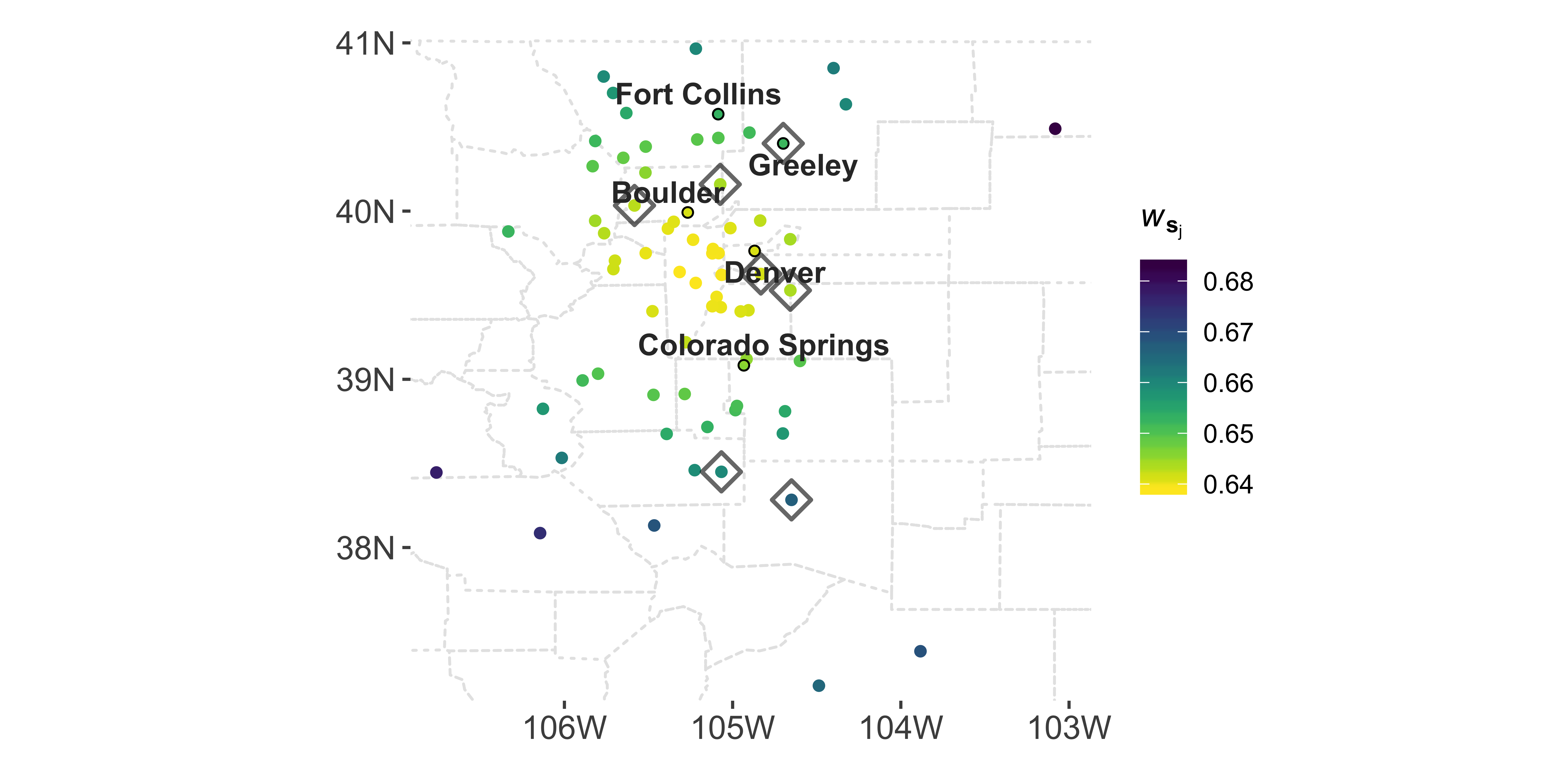} 

}

\caption[Spatial distribution of weights]{Spatial distribution of weights.  Weights are smaller for locations central to the spatial sampling pattern, where extremal dependence is more likely to impact data.  Cities used in the hold-out model comparison are marked by diamond outlines.}\label{fig:co_wts_deltas}
\end{figure}

\end{knitrout}

\begin{knitrout}
\definecolor{shadecolor}{rgb}{0.969, 0.969, 0.969}\color{fgcolor}\begin{figure}

{\centering \includegraphics[width=\maxwidth]{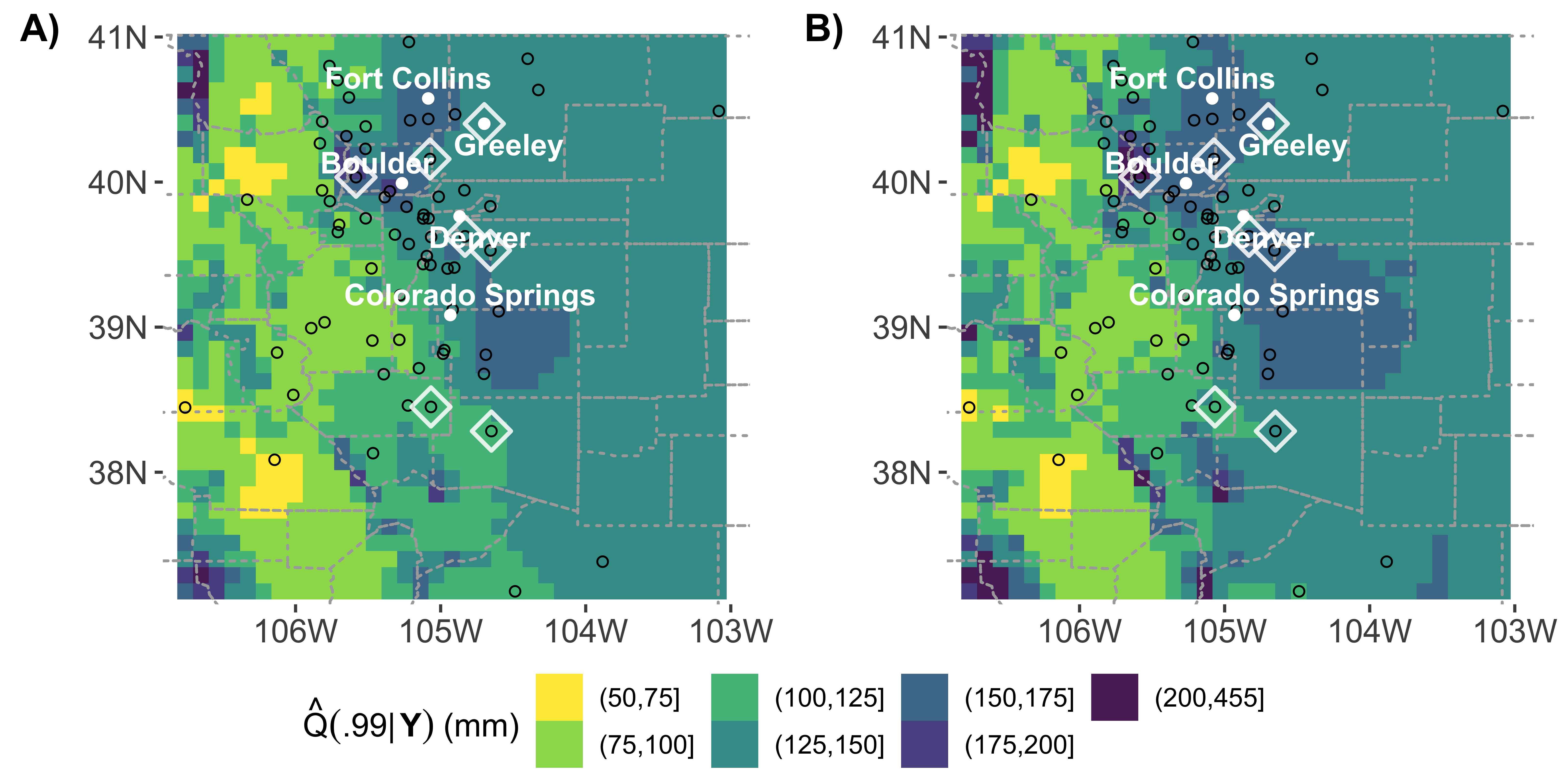} 

}

\caption{Spatially complete estimates $\hat Q\paren{\given{.99}{\v\eta\paren\vs}}$ of 100-year return levels for daily precipitation in Colorado's Front Range. Estimates are compared from the unweighted (A) and weighted (B) unweighted latent spatial extremes models.  The weighted estimates have increased smoothness and spatial range, and overall patterns that better match orographic features in Colorado.  The locations of the 71 stations whose data are analyzed are indicated by ($\circ$).  For reference, we include the names of several reference cities.   Cities used in the hold-out model comparison are marked by diamond outlines.}\label{fig:co_spatial_diffs}
\end{figure}

\end{knitrout}

\begin{table}[ht]
\centering
\caption{Comparison of log-scores for the weighted $\ell_{wtd}\paren{\vs_0}$ and unweighted models $\ell\paren{\vs_0}$ at holdout cities; the highest log-score is highlighted for each city.  The weighted likelihood model tends to have higher log-scores at holdout cities, suggesting better out-of-sample predictive performance in the targeted regions.  The low log-scores in the bottom row also suggest neither model is predictive of extreme precipitation in Pueblo.} 
\label{table:co_holdout_table}
\begin{tabular}{lrlrr}
  \toprule
Lat. & Lon. & City & $\ell_{wtd}\paren{\vs_0}$ & $\ell\paren{\vs_0}$ \\ 
  \midrule
40.4 & 104.7 & Greeley & \textbf{$-$224} & $-$225 \\ 
  40.2 & 105.1 & Longmont & \textbf{$-$502} & $-$508 \\ 
  40.0 & 105.6 & Nederland & \textbf{$-$411} & $-$415 \\ 
  39.6 & 104.8 & Aurora & \textbf{$-$303} & $-$307 \\ 
  39.5 & 104.7 & Parker & \textbf{$-$262} & $-$701 \\ 
  38.5 & 105.1 & Penrose & $-$198 & \textbf{$-$196} \\ 
  38.3 & 104.7 & Pueblo & \textbf{$-$349,848} & $-$1,779,826 \\ 
   \bottomrule
\end{tabular}
\end{table}

\section{Discussion}
\label{sec:discussion}

Estimating marginal return levels is an important step in planning for impacts
of natural hazards, especially those caused by precipitation.  Extreme
precipitation data have dependence, which makes estimation more complicated.
Models that explicitly account for dependence in the data have limited ability
to scale to large datasets, while models that assume conditional independence
in the data can scale well to large datasets, but do not account for dependence.
We develop a weighted likelihood that downweights observations from locations
central to the spatial sampling pattern in order to better estimate marginal
return levels.
We use the extremal coefficient in \autoref{eq:extcoeff} to construct weights
that downweight likelihood contributions from locations central to
the spatial sampling pattern, where observations tend to be most dependent.
Simulations confirm that the weighting scheme
improved the uncertainty quantification of the return level estimates in
situations when data have extremal dependence.  In application,
estimates from the weighted model better align with expected changes in
patterns of extreme precipitation caused by physical features, like mountains.

Since weighted likelihoods are computationally inexpensive, they may be a
useful technique to adopt in most settings where latent spatial extremes models
are employed.
Weighting adds $N$ additional multiplications per likelihood evaluation, whereas
alternatives like
penalization add $N$ additional function evaluations.
Penalization improves estimation for univariate extremes data at
a similar computational cost,
but its main purpose is to discourage models from exploring unrealistic or
undesirable regions of the parameter space, such as those with large shape
parameters $\xi\paren\vs$. As a result, penalized models underestimate
uncertainty almost as much as unweighted models.  Composite likelihood
corrections are more computationally
expensive \citep{Sharkey, Ribatet2012}. In practice, weighting encourages
borrowing strength across locations to improve estimates at each location.

Refining the likelihood weights \autoref{eq:wts} could further improve the
ability for marginal likelihoods to account for extremal dependence when
estimating marginal return levels.  For example, pairwise densities can be
derived for specific max-stable processes \citep[e.g.,][]{Padoan2010}.  Pairwise
densities explicitly model the dependence between pairs of observations, while
the extremal coefficient we use to build likelihood weights measures a summary
of extremal dependence instead.
Empirical Bayes--like procedures could be developed that use
likelihood weights based on pairwise densities to further improve the
performance of return level estimators.  While empirical Bayes procedures will
not fully account for estimation uncertainty (e.g., in estimating dependence
parameters in bivariate densities), the procedures may still provide a
fair compromise between computational complexity and accurate estimation of
uncertainty.

Weighting schemes are flexible, so may be extended to accommodate
complex issues in modeling and estimation outside extremes applications.
While we demonstrate the use of a weighted likelihood for latent spatial
extremes models, the theory we develop is more general.  The Fisher
information interpretation of weighted likelihoods also applies to all weighted
likelihoods.  Similarly, the limiting behaviors of likelihoods for independent
data or completely dependent data are largely based on copula theory for
arbitrary data, rather than extreme value theory.  Importantly, the
construction of the weighted likelihood \autoref{eq:wtdLik} can be adapted to
other statistical problems where marginal inference is of interest but
likelihoods are difficult to evaluate.  The construction we propose is based on
the idea that a computationally inexpensive measure of dependence between
observations can be used to develop a weighted likelihood that better quantifies
parameter uncertainty than related unweighted models.  The main challenge in
adapting our weighted likelihood to other applications is in identifying an
appropriate dependence measure that can be used to build likelihood weights.

\section*{Supplementary materials} Additional information and supporting
material for this article is available online at the journal's website.

\section*{Acknowledgements}

This material is based upon work supported by the National
Science Foundation under grant number AGS--1419558 (Hewitt and Hoeting) and
DMS--1243102 (Fix and Cooley). This research utilized the CSU ISTeC Cray HPC
System supported by NSF Grant CNS--0923386. This work utilized the RMACC Summit
supercomputer, which is supported by the National Science Foundation
(awards ACI-1532235 and ACI-1532236), the University of Colorado Boulder, and
Colorado State University. The Summit supercomputer is a joint effort of the
University of Colorado Boulder and Colorado State University.

We also express our gratitude to Emeric Thibaud and Mathieu Ribatet.  Dr.
Thibaud provided code to simulate Brown-Resnick processes, and Dr. Ribatet
provided a development
version of the \texttt{SpatialExtremes} package, written for the \texttt{R}
computing language, that implements a Gibbs sampler for the unweighted latent
spatial extremes model.

\bibliographystyle{apalike}
\bibliography{references}

\end{document}